\documentclass[
 reprint,
superscriptaddress,
 amsmath,amssymb,
 aps,
]{revtex4-2}

\usepackage{graphicx}
\usepackage{dcolumn}
\usepackage{bm}
\usepackage{braket}
\usepackage{orcidlink}

\usepackage{enumitem} 
\newcommand{\be}{\begin{equation}}
\newcommand{\ee}{\end{equation}}
\newcommand{\bea}{\begin{eqnarray}}
\newcommand{\eea}{\end{eqnarray}}
\DeclareUnicodeCharacter{039B}{$\Lambda$}

\newcommand{\W}{\Omega}

\usepackage{hyperref} 

\usepackage{physics}
\usepackage{float} 
\usepackage{listings}
\usepackage[usenames,dvipsnames,x11names]{xcolor}
\usepackage{tikz}

\begin{document}

\preprint{APS/123-QED}

\title{Fast Arbitrary Qutrit Gates for NV Centers in the Low-Field Regime}

\author{Alberto López--García\,\orcidlink{0009-0001-0023-9850}}%
\email{alberto.lopezg@upct.es}
\affiliation{\'Area de F\'isica Aplicada, Universidad Polit\'ecnica de Cartagena member of the European University of Technology EUT+, Cartagena E-30202, Spain}

\author{Marcel Morillas--Rozas\,\orcidlink{0009-0005-4570-1016}}
\affiliation{\'Area de F\'isica Aplicada, Universidad Polit\'ecnica de Cartagena member of the European University of Technology EUT+, Cartagena E-30202, Spain}

\author{Alberto Mayorgas\,\orcidlink{0000-0002-7071-2025}}
\affiliation{Escuela Polit\'ecnica Superior, Universidad CEU Fernando III, CEU Universities, Glorieta Cardenal Herrera Oria, 41930 Seville, Spain}

\author{Javier Cerrillo\,\orcidlink{0000-0001-8372-9953}}%
\email{javier.cerrillo@upct.es}
\affiliation{\'Area de F\'isica Aplicada, Universidad Polit\'ecnica de Cartagena member of the European University of Technology EUT+, Cartagena E-30202, Spain}


\begin{abstract}
The ground state of the negatively charged NV center forms a spin--1 manifold providing a versatile platform for sensing and information processing. Here we present a scheme for implementing fast arbitrary qutrit gates in the low--field regime using monochromatic microwave pulses of constant intensity tuned to the zero--field transition. By concatenating pulses with appropriate phases and durations, the NV-ERC scheme is extended from SU(2) operations in the double-quantum subspace to the full three--level structure. We show that arbitrary SU(3) operations can be decomposed into rotations in the double-quantum subspace together with effective implementations of the generators related to $\hat{\lambda}_5$ and $\hat{\lambda}_8$. We illustrate this decomposition with a use case: performing quantum state tomography of the complete three-level density matrix.
\end{abstract}


\maketitle


\section{\label{sec:level1}Introduction}

Nitrogen‑vacancy (NV) centers are multifunctional quantum systems that are conventionally controlled by addressing just two of their lowest energy levels \cite{Doherty2013,10.1116/5.0006785}. This selectivity is achieved by Zeeman-splitting  the $|\pm1\rangle$ levels through a bias magnetic field aligned with the NV axis and using microwave pulses resonant only with one of the two transitions $|0\rangle\leftrightarrow|\pm 1\rangle$. This approach implicitly sets an upper bound on the intensity of the pulses so as to suppress off-resonant excitation to the level that is not being addressed. When the pulse intensity becomes comparable to the Zeeman splitting, the two-level approximation breaks down, worsening coherence time and  sensitivity. To overcome this limitation, which is especially important at low magnetic fields, a three-level control  technique for NV centers based on an effective Raman coupling (NV ERC) was recently proposed that addresses all NV ground-state spin triplet levels simultaneously \cite{Cerrillo2021,Vetter2022}. This method enables precise control at low Zeeman splitting and allows the use of more intense pulses, expanding the achievable sensitivity range.

Three-level control is exploited in different physical-systems configurations, such as V-systems in semiconductor quantum dots \cite{PhysRevLett.95.187404}, atomic media and population trapping \cite{Liu2001,Kang:17,Ivander_2022,PhysRevA.74.063829,Gaoxiang_Li_2000}, and working fluids in heat engines \cite{Gelbwaser-Klimovsky2015}. There are also $\Lambda$-systems \cite{alma991081051849706532,Brion_2007,OU20084940} used in geometric and superadiabatic control in solid-state systems \cite{Yale2016,Rogers2016,Zhou2017}
, and cascade systems \cite{PhysRevA.47.519}. NV centers are an example of a V-system, where recent attempts of three-level control have been analyzed \cite{Cerrillo2021,Vetter2022,LopezGarcia2025FullQubitControlNV, PhysRevApplied.21.054011, Stark2018}.

Control protocol proposals are very relevant in high-dimensional quantum computation \cite{PhysRevLett.114.240401}, where there is an interest to extend algorithms to qudits \cite{10.3389/fphy.2020.589504,9069177,cozzolino}. Among all current research topics, the following stand out: trade-offs of quantum optimization using qutrits \cite{Bottrill_2025,ACAR2025129404,Goss2022}, quantum cryptographic schemes with qudits \cite{PhysRevLett.88.127902}, the optimized design of universal ternary gates \cite{7341452}, and resource-efficient digital quantum simulation \cite{Sawaya2020}. Examples of hardware for qudit quantum computation are superconducting systems \cite{KIKTENKO20151409,PhysRevX.11.021010} including transmons \cite{Cao_2024,PRXQuantum.4.030327}, a single photon \cite{singlephoton,PhysRevLett.119.180510}, the silicon-photonic quantum chip \cite{Chi2022}, trapped-ions \cite{PhysRevA.67.062313,Meth2025,Andrade_2022,Ringbauer2022, PhysRevLett.85.4458,cerrillo2010fast,cerrillo2018double}, and molecular magnets \cite{PhysRevLett.119.187702,Biard2021}. As a consequence, robust benchmarking methods have been developed to evaluate the performance of high-dimensional quantum computation hardware \cite{PhysRevLett.126.210504}. Additionally, the realm of qudit exploitation extends to other areas of quantum information including quantum communication \cite{cozzolino,Yu2025} or quantum simulation \cite{PhysRevA.110.062602}.

Generating qutrit gates introduces the challenge of moving from the usual picture in SU(2) to SU(3). The Pauli matrices generating the SU(2) algebra should be replaced by Gell-Mann matrices \cite{PhysRev.125.1067}. Hence, operations such as the commutation relations and the spin composition (tensor product) have to be redefined \cite{Barut,HallBook,GREINER}. In many-body systems, the inclusion of a third level increases the spectral complexity and Hilbert-space dimensionality \cite{nuestroPRE,PhysRevE.83.046208}. That is, a considerable number of avoided crossings emerge, a feature typical of chaotic dynamics, that interfere with the level clustering \cite{Stockmann1999,Casati,Meredith}. A large subset of quantum physics problems are based on qudits as principal components, such as vibrational modes \cite{PhysRevA.86.032508,Calixto_2012}, bosonic fundamental particles \cite{PhysRevB.40.546,PhysRevC.93.044302,iachello_arima_1987}, nuclear shell model \cite{Casati,Meredith,nuestroPRE}, and spin-s particles \cite{Levitt}. 

In this paper, we extend the NV‑ERC protocol to arbitrary SU(3) gates by expressing the corresponding unitary evolution in a form that naturally decomposes into elementary operations generated by a selected subset of the Gell‑Mann matrices. This subset is chosen to complete the set of achievable operations to the full SU(3) group. As a practical example, we apply the method to perform quantum state tomography of the full three‑level density matrix using constant-frequency pulses.

The paper is organized as follows. Section~\ref{sec:nverc} reviews the NV-ERC scheme in the double-quantum transition. Section~\ref{sec:su3_redef} introduces the equatorial state \(\ket{\xi(t)}\) and rewrites the unitary dynamics in a form suited for SU(3) formalism. Section~\ref{sec:su3_decomp} presents the corresponding SU(3) decomposition, while Secs.~\ref{sec:lambda8} and~\ref{sec:lambda5} develop the implementation of the \(\hat{\lambda}_{8}\) and \(\hat{\lambda}_{5}\) operations, respectively. Section~\ref{sec:application} contains the numerical validation and application, and Sec.~\ref{sec:conclusions} closes the paper.

\section{\label{sec:nverc}Review of NV-ERC}


The NV effective Raman coupling (NV-ERC) technique \cite{Cerrillo2021} enables precise initialization, manipulation, and readout of the double-quantum transition within the ground state of the NV center. The ground state manifold is described by a spin-\(1\) triplet. Its dynamics in the presence of an external magnetic field \(B\) and a microwave drive are governed by the Hamiltonian
\be
H=DS_z^2 + \mu B S_z + \Omega \cos \left( Dt - \alpha \right) S_x,
\label{eq:H}
\ee where \(D\) is the zero-field splitting, \(\mu B\) is the Zeeman splitting, and \(\Omega\) is the amplitude of the microwave driving with frequency \(D\) and an arbitrary phase \(\alpha\).

Denoting by \(\ket{0}\), \(\ket{+1}\), and \(\ket{-1}\) the  eigenstates of \(S_z\), it is useful to introduce the states
\bea
\ket{\pm}&=&\dfrac{1}{\sqrt{2}}\left(\ket{+1}\pm\ket{-1}\right).
\eea
Moving into the interaction picture with respect to \(H_0 = D S_z^2\), and within the rotating wave approximation (RWA), the Hamiltonian reduces to
\be
H'=\mu B \ket{-}\bra{+} +\frac{\W}{2} e^{i \alpha} \ket{+} \bra{0} + H.c.,
\label{eq:HRWA}
\ee
which represents an effective Raman coupling between \(|0\rangle\) and \(|-\rangle\) mediated by the state \(|+\rangle\).


The dynamics associated with Eq.~(\ref{eq:HRWA}) is analytically solvable in terms of the bright and dark states
\bea
\ket{B_\alpha}&=&\frac{1}{\bar\W}\left(\mu B \ket{-} + e^{-i\alpha}\frac{\W}{2}\ket{0}\right),\\
\ket{D_\alpha}&=&\frac{1}{\bar\W}\left(-\mu B \ket{0} + e^{i\alpha}\frac{\W}{2}\ket{-}\right),
\eea
where $\bar\W^2=\mu^2 B^2 + \W^2/4$. The corresponding evolution operator is
\begin{multline}
U(t, \alpha) = \cos\bar{\Omega} t \left(|B_\alpha\rangle\langle B_\alpha| + |+\rangle\langle+|\right) \\
- i \sin\bar{\Omega} t \left(|B_\alpha\rangle\langle+| + |+\rangle\langle B_\alpha|\right) + |D_\alpha\rangle\langle D_\alpha|.
\label{eq:Ut}
\end{multline}
where the dark state \(|D_\alpha\rangle\) remains decoupled from the dynamics, while \(|B_\alpha\rangle\) and \(|+\rangle\) undergo coherent Rabi oscillations of frequency $\bar\W$.


NV-ERC relies on a unique feature of Eq.~(\ref{eq:Ut}): there exists a  characteristic time \(\bar{T}'\) such that \(U(\bar{T}',\alpha)\) maps \(\ket{0}\) to the state
\be
\ket{\phi}= \dfrac{e^{-i\frac{\phi}{2}}\ket{-1}-e^{i\frac{\phi}{2}}\ket{+1}}{\sqrt{2}},
\label{phi}
\ee
where $\phi=2\arccos(2\mu B/\Omega)$. This state lies on the equator of the double-quantum Bloch sphere spanned by the states \(|+1\rangle\) and \(|-1\rangle\). The characteristic time \(\bar{T}'\) is given by
\be
\bar{T}' = \dfrac{\arccos\left(-\frac{4\mu^2B^2}{\Omega^2}\right)}{\bar{\Omega}},
\label{T_prime}
\ee  
and is valid for any value \(\Omega \ge 2 \mu B\), which means that the protocol can be implemented in both the fast-pulse and low-field regimes \cite{Cerrillo2021, Vetter2022}.\\

Aided by this realization, it was possible to express arbitrary qubit gates within the double-quantum transition \cite{LopezGarcia2025FullQubitControlNV} by the concatenation of pulses
\bea
U(\bar T', \alpha)
&=& U^\dagger(\bar T'', \alpha) \\
\nonumber &=& e^{i\alpha}\ket{\phi}\bra{0}
  + e^{-i\alpha}\ket{0}\bra{-\phi}
  - \ket{\pi+\phi}\bra{\pi-\phi}
\label{eq:Ut1}
\eea
where \(\bar{T}'' = 2\pi / \bar{\Omega} - \bar{T}'\). Combining a pulse of duration \(\bar{T}'\) and phase \( \alpha \) with a pulse of duration \(\bar{T}''\) and phase \(\alpha + \theta\) yields a rotation of angle \( \theta \) around the \( -\phi\) axis of the double-quantum Bloch sphere, while the ground state acquires a phase \( \theta\):
\begin{align}
U_{\text{pair}}(\phi, \theta)
&= U(\bar T'', \alpha + \theta)\, U(\bar T', \alpha) \nonumber \\
&= e^{-i\theta}\ket{0}\bra{0}
  + e^{i\theta}\ket{-\phi}\bra{-\phi}
  + \ket{\pi-\phi}\bra{\pi-\phi}.
\label{eq:Upair}
\end{align}
As a result, any single-qubit control operation in the double-quantum transition may be implemented using MW pulses at frequency $D$ of constant intensity. 

\section{\label{sec:su3_redef}Redefining the unitary operation for SU(3) symmetry}

\begin{figure}[t]
    \centering
    \includegraphics[width=0.965\columnwidth]{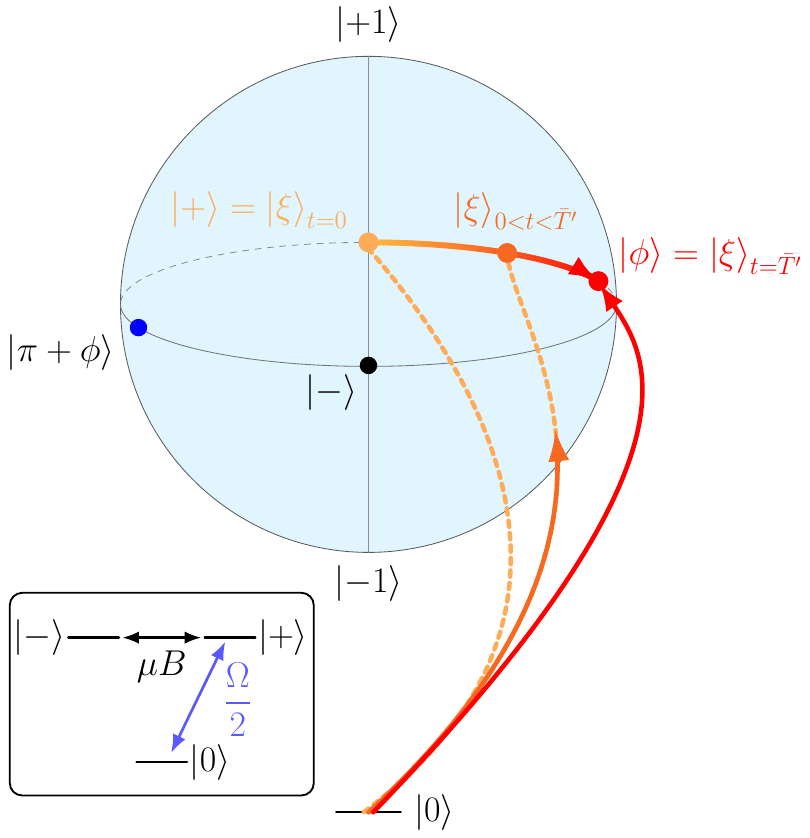}
    \caption{Evolution generated by \(U(t,\alpha)\) on the double-quantum Bloch sphere. The equatorial path connects \(\ket{+}=\ket{\xi(0)}\) to \(\ket{\phi}=\ket{\xi( T')}\), passing through \(\ket{\xi(0<t<\bar T')}\). The curved arrows show the transition from \(\ket{0}\) to the corresponding equatorial states. The inset shows the effective Raman coupling in the three-level scheme, with couplings \(\mu B\) and \(\Omega/2\).}
    \label{fig:xi_scheme}
\end{figure}

\begin{figure}[t]
    \centering
    \includegraphics[width=\columnwidth]{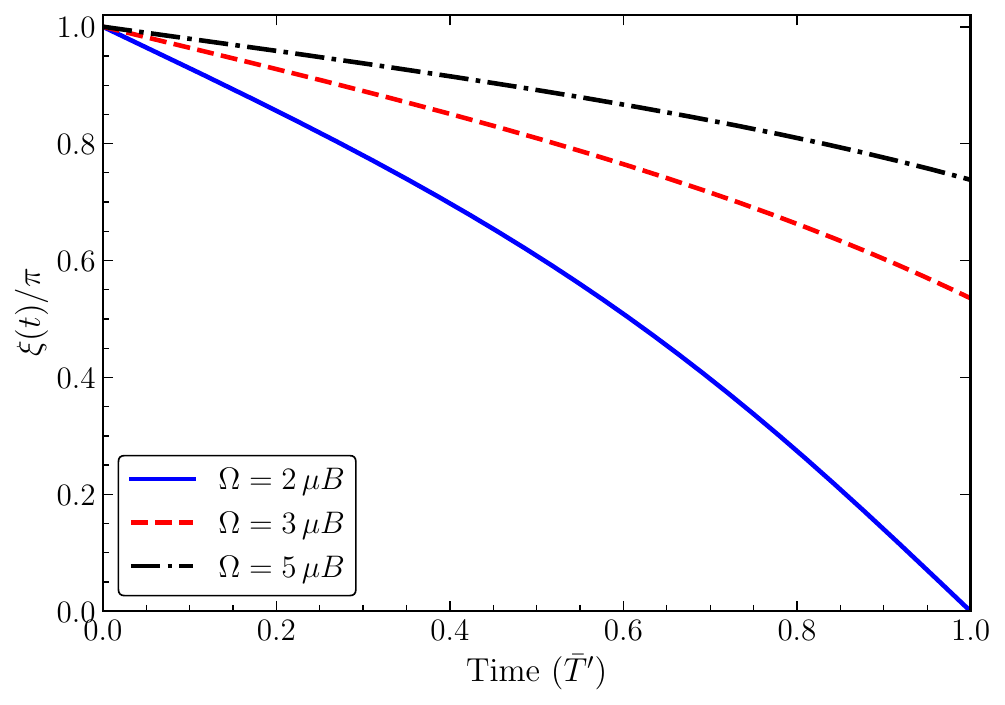}
    \caption{Evolution of the angle $\xi(t)$ for three driving amplitudes, $\Omega=2\mu B$, $\Omega=3\mu B$, and $\Omega=5\mu B$. Time is expressed in units of $\bar{T}'$, and the angle is shown in units of $\pi$.}
    \label{fig:xi_dynamics}
\end{figure}

In order to extend arbitrary gates to the full Hilbert space, it is convenient to rewrite Eq.~(\ref{eq:Ut}) in terms of states located on the equator of the double-quantum Bloch sphere. To this end, we note that, for arbitrary time \(t\), the state $U(t, \alpha)\ket{0}$ remains in the subspace spanned by $\ket{0}$ and an equatorial state of the double-quantum transition, which we denote by $\ket{\xi(t)}$:
\be
\ket{\xi(t)} =
\frac{1}{\eta}\left[
\mu B(\cos\bar{\Omega}t -1)\ket{-}
- i\bar{\Omega}\sin\bar{\Omega}t\ket{+}
\right],
\ee
where $\eta^2 = [ \mu B(\cos(\bar{\Omega } t) -1)]{^{2}} +[\bar{\Omega }\sin(\bar{\Omega } t)]^{2}$. It subtends an angle $\xi (t)$ with state $\ket-$ as defined implicitly by:
\be
\braket{-}{\xi(t)}
=
\frac{1}{\eta}\left[\mu B(1-\cos\bar{\Omega}t)\right]
=
\cos\!\left[\frac{\xi(t)}{2}\right].
\ee
Thus, the angle evolves continuously along the equator of the double-quantum transition, starting from $\xi(0) = \pi$ (which corresponds to the state $\ket{+}$ up to a global phase) and reaching $\xi(\bar{T}') = \phi$ at the characteristic time \(\bar{T}'\) (see Appendix B for details). This intermediate state dynamics is illustrated schematically in Fig.~\ref{fig:xi_scheme}, and its dependence on the ratio \(\Omega/\mu B\) is shown in Fig.~\ref{fig:xi_dynamics}.

To facilitate the SU(3) analysis below, it is convenient to rewrite the unitary operator in Eq. \eqref{eq:Ut} in terms of equatorial states of the double-quantum subspace together with \(\ket{0}\). In particular, using the states \( \ket{\xi}\), \(\ket{-\xi}\), \(\ket{\pi+\xi}\), and \(\ket{\pi -\xi}\), Eq.~(\ref{eq:Ut}) can be expressed as
%
\begin{multline}
U(t,\alpha)
=
A(t)\big(\ketbra{0}{0}-\ketbra{\xi}{-\xi}\big)
\\
-B(t)\big(e^{i\alpha}\ketbra{\xi}{0}+\ketbra{0}{-\xi}e^{-i\alpha}\big)
-\ketbra{\pi+\xi}{\pi-\xi},
\label{eq:U_xi}
\end{multline}
where \(\xi \equiv \xi(t)\),
\be
A(t)=
\cos(\bar{\Omega} t)\left(\frac{\Omega}{2\bar{\Omega}}\right)^2
+
\left(\frac{\mu B}{\bar{\Omega}}\right)^2,
\ee
and
\be
B(t)=\sqrt{1-A^2(t)}.
\ee
A detailed derivation is provided in Appendix C.

\section{\label{sec:su3_decomp}Decomposition of SU(3) operations}

In order to relate the new expression of $U(t,\alpha)$ to arbitrary qutrit gates, we now turn to a decomposition of a general SU(3) operator in terms of a subset of the Gell-Mann matrices \cite{Byrd_1998}:
\be
S =
e^{i\hat{\lambda}_{3}\delta}
e^{i\hat{\lambda}_{2}\epsilon}
e^{i\hat{\lambda}_{3}\zeta}
e^{i\hat{\lambda}_{5}\theta_{5}}
e^{i\hat{\lambda}_{3}a}
e^{i\hat{\lambda}_{2}b}
e^{i\hat{\lambda}_{3}c}
e^{i\hat{\lambda}_{8}\phi_{8}},
\ee
where \(\delta\), \(\epsilon\), \(\zeta\), \(\theta_{5}\), \(a\), \(b\), \(c\), and \(\phi_{8}\) are the necessary rotation angles specifying the decomposition. 

In the basis \(\{\ket{+},\ket{-},\ket{0}\}\), the operators generated by \(\hat{\lambda}_{2}\) and \(\hat{\lambda}_{3}\) act within the double-quantum subspace, spanned by \(\{\ket{+},\ket{-}\}\). Therefore, the previous expression can be rearranged as
\be
S = U_{DQ}^{A}\, e^{i\theta_{5}\hat{\lambda}_{5}}\, U_{DQ}^{B}\, e^{i\phi_{8}\hat{\lambda}_{8}},
\label{eq:su3_grouped}
\ee
where
\be
U_{DQ}^{A}=e^{i\hat{\lambda}_{3}\delta}e^{i\hat{\lambda}_{2}\epsilon}e^{i\hat{\lambda}_{3}\zeta},
\qquad
U_{DQ}^{B}=e^{i\hat{\lambda}_{3}a}e^{i\hat{\lambda}_{2}b}e^{i\hat{\lambda}_{3}c},
\ee
are arbitrary SU(2) rotations in the double-quantum transition. 

Equation~(\ref{eq:su3_grouped}) shows that universal single-qutrit control can be achieved once three elements are defined: arbitrary rotations within the double-quantum subspace, a gate generated by \(\hat{\lambda}_{5}\), and a gate generated by \(\hat{\lambda}_{8}\). Since NV-ERC already provides arbitrary control in the double-quantum transition \cite{LopezGarcia2025FullQubitControlNV}, the remaining task is to implement \(\hat{\lambda}_{5}\) and \(\hat{\lambda}_{8}\) operations within the same framework.

In the following sections, we show how both generators can be realized using monochromatic pulses of constant frequency defined in \cite{LopezGarcia2025FullQubitControlNV}.

\section{\label{sec:lambda8}lambda 8}

The Gell-Mann matrix
\[
\hat{\lambda }_{8} = \dfrac{1}{\sqrt{3}}
\begin{pmatrix}
1 & 0 & 0\\
0 & 1 & 0\\
0 & 0 & -2
\end{pmatrix}
\]
introduces a phase difference between the double-quantum subspace and the ground state \(\ket{0}\). 
\bea
\nonumber e^{i\phi_{8}\hat{\lambda}_{8}}
&=&
e^{i\phi_{8}/\sqrt{3}}
\big(\ketbra{+}{+}+\ketbra{-}{-}\big)\\
&&+
e^{-i2\phi_{8}/\sqrt{3}}\ketbra{0}{0}.
\label{eq:lambda8_expanded}
\eea
In this subsection it is convenient to express Eq.\eqref{eq:lambda8_expanded} in terms of states \(\ket{\phi}\) and \(\ket{\pi+\phi}\):
\bea
\nonumber
e^{i\phi_{8}\hat{\lambda}_{8}}
&=&
e^{i\phi_{8}/\sqrt{3}}
\big(\ketbra{\phi}{\phi}+\ketbra{\pi+\phi}{\pi+\phi}\big)
\\&&+
e^{-i2\phi_{8}/\sqrt{3}}\ketbra{0}{0}.
\label{eq:lambda8_expanded2}
\eea
In the NV-ERC approach, any rotation of angle \(\theta\) applied to the double-quantum transition adds a phase \(-\theta\) to the state \(\ket{0}\). With \(\hat{\lambda}_{8}\) the concatenated pair Eq.\eqref{eq:Upair} can be expressed as
\be
U_{\mathrm{pair}}(\phi,\theta)
=
e^{-i\theta/3}e^{i\theta\hat{\lambda}_{8}/\sqrt 3}
R(\phi,\theta),
\label{eq:lambda8_pair_def}
\ee
where \(\theta\) denotes the relative phase shift between the pair of pulses and we define the effective rotation 
\be
R(\phi,\theta)
=
e^{i\theta}\ketbra{\phi}{\phi}
+
\ketbra{\pi+\phi}{\pi+\phi}
+\ketbra{0}{0}.
\label{eq:lambda8_pair}
\ee
within the double-quantum subspace. Thus, each pulse pair produces the phase shift \(e^{-i\theta}\) on \(\ket{0}\), together with the effective double-quantum rotation \(R(\phi,\theta)\).

Now, we combine
\be
U_{\mathrm{pair}}(\phi,\theta)U_{\mathrm{pair}}(\pi+\phi,\theta)
=
e^{i\phi_{8}\hat{\lambda}_{8}},
\ee
where \(U_{\mathrm{pair}}(\pi+\phi,\theta)\) is defined analogously to
Eq.~(\ref{eq:lambda8_pair_def}) with
\[
R(\pi+\phi,\theta)
=
\ketbra{\phi}{\phi}
+
e^{i\theta}\ketbra{\pi+\phi}{\pi+\phi}
+\ketbra{0}{0}.
\]
and
\be
\phi_8=\sqrt{3}\,\theta.
\label{eq:phi8_theta_relation}
\ee

\section{\label{sec:lambda5}lambda 5}

\begin{figure}[t]
    \centering
    \includegraphics[width=\columnwidth]{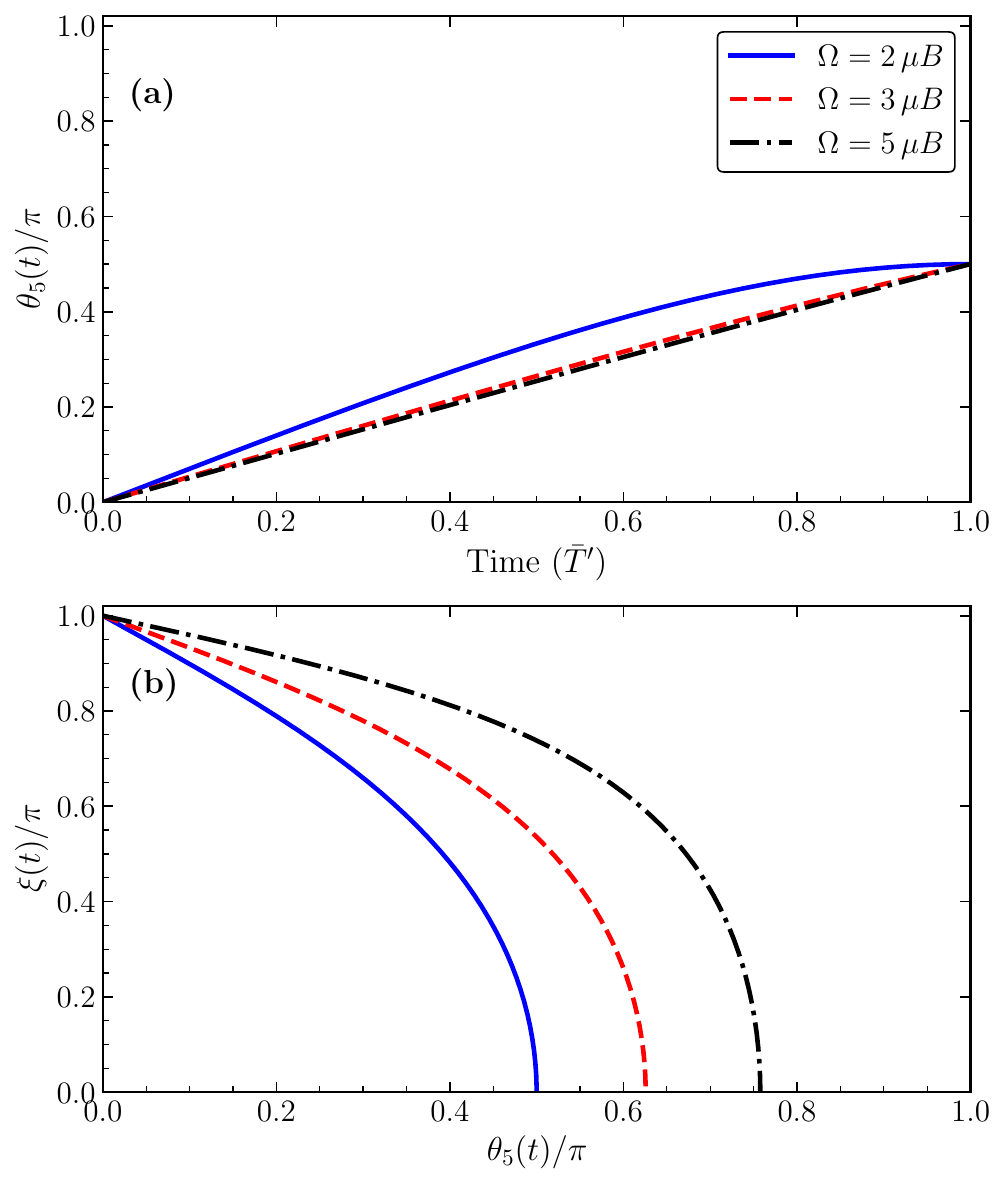}
    \caption{\textbf{(a)} Evolution of the effective rotation angle \(\theta_{5}(t)\) for \(\Omega=2\mu B\), \(\Omega=3\mu B\), and \(\Omega=5\mu B\), with time given in units of \(\bar T'\). \textbf{(b)} Parametric relation between \(\xi(t)\) and \(\theta_{5}(t)\) for the same driving amplitudes. Angles are shown in units of \(\pi\).}
    \label{fig:theta5_xi}
\end{figure}

The Gell-Mann matrix
\[
\hat{\lambda }_{5} =
\begin{pmatrix}
0 & 0 & -i\\
0 & 0 & 0\\
i & 0 & 0
\end{pmatrix}
\]
generates a transition between the states \(\ket{0}\) and \(\ket{+}\) in the basis \(\{\ket{+},\ket{-},\ket{0}\}\). Its exponential can be written as
\begin{multline}
e^{i\theta_{5}\hat{\lambda}_{5}}
=
\cos\theta_{5}\big(\ketbra{0}{0}+\ketbra{+}{+}\big)
\\
+\sin\theta_{5}\big(\ketbra{+}{0}-\ketbra{0}{+}\big)
+\ketbra{-}{-}.
\label{eq:lambda5_expanded}
\end{multline}
Comparing to Eq.~(\ref{eq:U_xi}), we see
\be
U(t,\alpha)=
e^{i\frac{\alpha}{\sqrt3}\hat{\lambda}_8}
e^{i\xi \hat{\lambda}_2}
e^{i\hat{\lambda}_5\theta_5}
e^{i\xi \hat{\lambda}_2}
e^{-i\frac{\alpha}{\sqrt3}\hat{\lambda}_8},
\label{eq:lambda5_decomp}
\ee
with
\be
\cos\theta_{5}(t)=A(t),
\qquad
\sin\theta_{5}(t)=B(t).
\label{eq:theta5_def}
\ee
Equivalently,
\be
\cos\theta_{5}(t)=\bra{0}U(t,\pi)\ket{0}.
\ee
With a judicious combination of $U_{\text{pair}}$ and $U(t)$, it is possible to cancel the unwanted rotations generated by $\hat{\lambda}_{2}$ and $\hat{\lambda}_{8}$ before and after the targeted $\hat{\lambda}_{5}$ operation. The resulting effective dynamics isolates the target generator, so that the net evolution corresponds to a controlled rotation in the $\hat{\lambda}_{5}$ subspace. The dependence of $\theta_{5}(t)$ on the pulse duration, together with its relation to $\xi(t)$, is illustrated in Fig.~\ref{fig:theta5_xi}.

\section{Quantum State Tomography of a Qutrit with NV centers}
\label{sec:application}
 In this section, we show that we can use the SU(3) protocol presented in this work to perform full quantum state tomography of the density matrix with pulses of constant frequency. We describe the set of operations and measurements required under the constraint that only the population of the state $\ket{0}$ can be directly measured, as is standard in fluorescence readout of NV centers in diamond.

We consider the basis $\lbrace\ket{+},\ket{-},\ket{0}\rbrace$. In this basis the density matrix of the NV center is
\begin{equation}
    \rho = \mqty(p_{+}&c_{+,-}&c_{+,0}\\ c^*_{+,-} & p_-&c_{-,0}\\ c^*_{+,0}&c^*_{-,0}&p_{0}),
\end{equation}
where $p_i=\mel{i}{\rho}{i}\in\mathbb{R}$ denotes the population in state $\ket{i}$ and
$c_{i,j}=\mel{i}{\rho}{j}\in\mathbb{C}$ represents the coherence between states $\ket{i}$ and $\ket{j}$. Because the density operator is Hermitian, it is sufficient to reconstruct either the upper or the lower triangular part of the matrix together with the diagonal elements. Furthermore, since $\text{Tr}(\rho) = 1$, it is not necessary to measure the populations in all of the states. In particular, measuring $p_0$ and $p_-$ is sufficient, since the remaining population can be obtained as $p_+ = 1 - p_0 - p_-$. Consequently, instead of determining all nine matrix elements, the full state can be reconstructed by measuring five independent quantities. Since the coherences are complex numbers, both their real and imaginary parts must be determined separately, amounting to eight real parameters.

As mentioned above, $p_0$ can be directly obtained via the fluorescence of the NV center in diamond. To determine $p_-$, the measurement has to be preceded by a  rotation $e^{-i\phi\hat{\lambda}_3}$ mapping $\ket{-}$ to $\ket{\phi}$ followed by the unitary $U(\bar{T}'')$ mapping $\ket{\phi}$ to $\ket{0}$.

Estimating the coherences requires a more elaborate procedure since both their real and imaginary parts must be extracted. We begin with the coherence $c_{+,-}$. This coherence lies entirely within the double-quantum transition and can be accessed using $\hat{\lambda}_2$ and $\hat{\lambda}_3$ rotations. To obtain the real part of $c_{+,-}$:
\begin{enumerate}
    \item Apply the rotation $e^{i\hat{\lambda}_2(\pi /2-\phi)}$, mapping the state $\frac{1}{\sqrt{2}}(\ket{+1} + i\ket{-1})$ to $\ket{\phi}$.
    \item Apply the pulse $U(\bar{T}'')$ mapping $\ket{\phi}$ to $\ket{0}$ and measure the fluorescence.
\end{enumerate}
The imaginary part of the coherence is measured by preceding the previous sequence with $e^{i\hat{\lambda}_3\pi /2}$, mapping $\ket{+1}$ to $\frac{1}{\sqrt{2}}(\ket{+1} + i\ket{-1})$.

To determine $c_{0,+}$ and $c_{0,-}$, rotations $\hat{\lambda}_5$ and $\hat{\lambda}_8$ are required. To measure the real part of $c_{0,+}$, the protocol is as follows:
\begin{enumerate}
\item Apply the unitary rotation generated by $\lambda_5$ with angle $\theta_5=\pi/4$, to map $\frac{1}{\sqrt{2}}(\ket{0} - \ket{+})$ to $\ket{0}$:
    $$U_{\hat{\lambda}_5}\left(\frac{\pi}{4}\right) = \ket{0}\frac{\bra{0} - \bra{+}}{\sqrt{2}} + \ket{+}\frac{\bra{0} + \bra{+}}{\sqrt{2}} + \ketbra{-}$$
    \item Measure the fluorescence.
\end{enumerate}
For the imaginary part of $c_{0,+}$, precede the previous procedure with $U_{\hat{\lambda}_8} (\phi_8)= e^{i\phi_8\hat{\lambda}_8}$ such that $(\ket{0} + i\ket{+})/\sqrt{2}$ is mapped to $(\ket{0} + \ket{+})/\sqrt{2}$.

The coherence $c_{0,-}$ can be obtained by preceding the procedure for $c_{0,+}$ with a $\hat{\lambda}_2$ rotation that transfers $\ket{-}$ to $\ket{+}$. This completes all the measurements necessary for qutrit quantum state tomography.

\section{\label{sec:conclusions}Conclusions}

In this work, we have extended the NV-ERC framework from arbitrary SU(2) control of the double-quantum transition to full control of the three-level structure of the NV center. Starting from the effective Raman description and dynamics, we have shown that the associated unitary evolution can be reformulated in a natural way within the SU(3) structure of the spin-1 triplet system.

Through this reformulation, we have identified that the decomposition of arbitrary qutrit operations can be reduced to three basic operations: arbitrary rotations in the double-quantum transition together with the operations associated with the generators \(\hat{\lambda}_{5}\) and \(\hat{\lambda}_{8}\). Since NV-ERC already provides full SU(2) control of the double-quantum subspace, the problem is reduced to the effective implementation of the remaining \(\hat{\lambda}_{5}\) and \(\hat{\lambda}_{8}\) operations within the same NV-ERC framework, using monochromatic pulses of constant amplitude, specific phases, and suitable pulse concatenations.

This solution provides a route toward the implementation of fast arbitrary qutrit gates in NV centers in an intuitive way and under relatively simple experimental conditions, without the need for strong magnetic fields or complex pulse sequences and combinations. In this sense, the proposed scheme preserves the main advantage of the NV-ERC protocol while extending its scope to full control of the three-level system.

As a relevant application, we have shown that the protocol enables complete quantum state tomography of a qutrit under the constraint that only the ground state \(\ket{0}\) can be directly measured. Therefore, the scheme is not limited to coherent spin control, but also provides a practical route toward state characterization and readout.

Overall, these results establish NV-ERC as a flexible framework for the control of NV centers. Beyond the applications discussed here, the scheme may serve as a foundation for future developments in quantum sensing and quantum information, as well as for the extension of this protocol to other architectures with analogous structure.

\section{Acknowledgments}

All authors acknowledge support from grant CNS2023-144994 funded by MICIU/AEI/10.13039/201100011033 and by ``ERDF/EU'', as well as support from European Union project C-QuENS (Grant No. 101135359).
\newpage

\bibliography{NVQ3} 
\bigskip  

\newpage

\end{document}